\documentclass[11pt]{article}

\usepackage{latexsym, amsmath, amssymb}
\usepackage{psfrag}
\usepackage{graphicx}
\usepackage{mathrsfs}
\usepackage{bbm}
\usepackage{psfig}

\usepackage{latexsym, amsmath, amssymb}
\usepackage{psfrag}
\usepackage{psfig}
\usepackage{subfigure}
\usepackage{graphicx}
\usepackage{mathrsfs}
\usepackage{bbm}

\newcommand{\quarter}{{1\over4}}
\newcommand{\be}{\begin{eqnarray*}}
\newcommand{\ee}{\end{eqnarray*}}
\newcommand{\bea}{\begin{eqnarray}}
\newcommand{\eea}{\end{eqnarray}}
\newcommand{\ms}{\medskip}

\newcommand{\qbps}{\quarter\mathrm{-BPS}}

\newcommand{\p}{\partial}
\newcommand{\ds}{{\p \over \p \sigma}}
\newcommand{\tr}{\mathrm{tr}}
\newcommand{\Tr}{\mathrm{Tr}}
\newcommand{\Title}[1]{
  ~\vspace{12mm}
  \begin{center}
    \Large\bf #1
  \end{center}
  }
\newcommand{\Author}[1]{
  \begin{center}
    \large #1
  \end{center}
  }
\newcommand{\Institution}[1]{
  ~\vspace{-26pt}
  \begin{center}
   #1
  \end{center}
  }
\newcommand{\Email}[1]{{\tt #1}}

\newcommand{\History}[1]{
\centerline{\small #1}
}

\begin{document} \setlength{\unitlength}{1mm}

%\begin{titlepage}
\thispagestyle{empty}
\rightline{\tt hep-th/0510127}
\Title{Bloch Waves and Fuzzy Cylinders:\\
 $\frac{1}{4}$-BPS Solutions of Matrix Theory}
\bigskip
\Author{Peter G. Shepard}
\Institution{
\it{Center for Theoretical Physics and Department of Physics, University of California\\
Berkeley, CA 94720-7300, USA\\
and\\
Theoretical Physics Group, Lawrence Berkeley National 
Laboratory\\
Berkeley, CA 94720-8162, USA}}
\begin{center}
\Email{pgs@socrates.berkeley.edu}
\ms\ms
\History{October 16$^{th}$, 2005}
\end{center}
\ms\ms

\begin{abstract} 
In this note, we present a broad class of $\qbps$ solutions to matrix theory, corresponding to non-commutative cylinders of arbitrary cross-sectional profile in $\mathbf{R}^8$.  The solutions provide a microscopic description of a general supertube configuration.  Taking advantage of an analogy between a compact matrix dimension and the Hamiltonian of a 1-dimensional crystal, we use a Bloch wave basis to diagonalize the transverse matrices, finding a distribution of eigenvalues which smoothly trace the profile curve as the Bloch wave number is varied.
\end{abstract}

\newpage

\section{Introduction}

The dimensional reduction of ten-dimensional U($N$) super Yang Mills theory to 0+1 dimensions has been seen to contain a remarkable structure.  Since its original consideration in the 80's \cite{Marty}, it has emerged as the worldline theory of a collection of D0-branes, and more recently as a non-perturbative definition of M-theory \cite{BFSS}. 

Central to its power is the non-commutative geometry of its 9 adjoint-valued scalars, which are capable of describing not only point-like objects, but the extended branes of string theory and M-theory \cite{BSS, Matrix} as various ``fuzzy'' objects.  However, understanding how the matrix model degrees of freedom assemble themselves into such objects remains mysterious.\footnote{For another recent view on this issue, see \cite{Lexi}.}

In this note, we exploit an analogy between a compact matrix dimension and the Hamiltonian of a one-dimensional crystal, to build a broad class of $\qbps$ non-commutative cylinder solutions.  The analogy motivates a Bloch-wave basis which diagonalizes the transverse matrices, yielding distributions of eigenvalues familiar from solid-state band theory, which appear as the shape of the cylinder profile traced out by a continuum of eigenvalues in $\mathbf{R}^8$.

The solutions below provide a microscopic matrix model description of the general configuration of supertubes.\footnote{See \cite{BL,LRS,BOS}, for previous constructions in terms of operator algebras.  See also \cite{HO,BK,BO} for related results.}  Supertubes \cite{Supertubes} are bound states of D0 branes, fundamental string charge and angular momentum, which by the Myers effect \cite{Myers} puff-up into a tubular D2, supported by the Poynting vector of crossed electric and magnetic Born-Infeld fluxes.  

By demonstrating how the curves reemerge from the matrix eigenvalues, we learn a great deal about these non-commutative objects and how matrix degrees of freedom reorganize themselves into an extended object.

Though everything below is equally applicable to M(atrix)-theory, will use the language of Type IIa.

\section{The Matrix Theory}
We consider the matrix theory Hamiltonian
\be
H= \mathrm{tr}(D_0X_I^2 -[X_I,X_J]^2)+fermions
\ee
where $I,J=1\dots9$, and $D_0 X=\p_t+i[A_0,X]$. The gauge field $A_0$ is non-dynamical, acting as an auxiliary field to enforce the Gauss-law constraint
\bea
\sum_{I=1}^9 [X_I,D_0X_I]=0. \label{Gauss}
\eea
To derive BPS conditions, we pick out a direction $Z=X_9$, and rewrite the bosonic terms as\footnote{This is a generalization of \cite{BL}, with two errors corrected.} 
\bea
H &=& \tr\left(\sum_{i=1}^8\left(D_0 X_i + i[Z,X_i]\right)^2+ D_0Z^2 - \sum_{i,j=1}^8[X_i,X_j]^2 + C\right) \geq \tr(C),\label{H}\\
C &=& -i\sum_{i=1}^8\{D_0 X_i,[Z,X_i]\}. \label{C}
\eea
The BPS solutions are those that saturate the inequality in (\ref{H}), requiring the positive-definite terms to vanish\footnote{Recall that $[A,B]$ is negative definite.}
\bea
D_0 X_i + i[Z,X_i]&=&0 \nonumber \\
D_0 Z &=&0 \label{BPS} \\
\left[X_i,X_j\right]&=&0.\nonumber
\eea
As discussed in \cite{BSS,BL} such solutions preserve 8 supercharges.

\section{Matrix Compactification}
Since our interest is in translationally-invariant objects, we will consider solutions wrapped on a circle of circumference $L$, with the understanding that non-compact solutions are recovered by going to the covering space.  Taylor \cite{Taylor} showed how to compactify a matrix dimension by quotienting the covering space by discrete translations $\mathbf{Z}$.  $N$ D0 branes in a compact dimension are described by those states of an array of $N\times\infty$ D0's which transform under the generator $T$ of $\mathbf{Z}$ as
\be
T^\dagger Z T &=& Z+L\\
T^\dagger X_i T &=& X_i.  
\ee
Such matrices are infinite matrices of matrices, taking the form
\be
Z &=&
\left( \begin{array}{cccccccc}
\dots & Z(2)^\dagger & Z(1)^\dagger & Z(0)-L & Z(1) & Z(2) & \dots &\dots\\
\dots & \dots & Z(2)^\dagger & Z(1)^\dagger & Z(0) & Z(1) & Z(2) &\dots\\
\dots & \dots & \dots & Z(2)^\dagger & Z(1)^\dagger & Z(0)+L & Z(1)& Z(2)\\
\dots & \dots & \dots & \dots & Z(2)^\dagger & Z(1)^\dagger & Z(0)+2L&Z(1)\\
\end{array}\right),
\ee
or
\bea
Z_{mn} &=& mL\delta_{mn} + \sum_{l=-\infty}^\infty Z(l) \delta_{m,n+l}, \label{Z} \\
X^i_{mn} &=& \sum_{l=-\infty}^\infty X^i(l) \delta_{m,n+l}, \label{X}\\
A^0_{mn} &=& \sum_{l=-\infty}^\infty A^0(l) \delta_{m,n+l}. \label{A}
\eea
The indeces $m,n=-\infty \dots \infty$ label the copy of $S^1$ in the covering space, and the objects $X^\mu(l)$ are $N\times N$ matrices.\footnote{We will use $X^\mu$ as shorthand for $(A_0,X_i,Z)$, and will frequently use an upper index $X^i$ instead of $X_i$ to avoid crowding with the matrix indeces, with no metric sign-flip to be inferred.}  Hermiticity requires $X^\mu(-l)=X^\mu(l)^\dagger$.  The diagonal $Lm$ term in $Z$ indicates the shifted positions of the $m^{th}$ image of the D0's in the covering space, while the elements $X^\mu(l)$ represent strings connecting a D0 and its $l^{th}$ image.

It is convenient to introduce a basis of shift-invariant matrices consisting of a $\mathbf{1}_{N\times N}$ in each row $l$ spaces to the right of the diagonal, denoted by the symbol
\be
(e^{il\sigma})_{mn} \equiv \delta_{m,n-l},
\ee
and the ``derivative'' operator $\ds$ defined by $[\ds,e^{il\sigma}]=ile^{il\sigma}$, represented by
\be
\left(\ds\right)_{mn} = im\delta_{mn}.
\ee
Equations (\ref{Z} - \ref{A}) now take the form
\be
Z &=& -iL\ds + L A^\sigma(\sigma),\\
X^i &=& X^i(\sigma),\\
A^0 &=& A^0(\sigma),
\ee
where
\be
X^i(\sigma)= \sum_{l=-\infty}^\infty X^i(l) e^{il\sigma},
\ee
and similarly for $A^0$ and $A^\sigma=Z(\sigma)/L$.  

In fact, this representation makes explicit the T-duality between D0's and D1's on $S^1$.  The operator $e^{i\sigma}$ represents a coordinate along the D1's wrapping $S^1$, with the fields $A^{0,\sigma}$ and $X^i(\sigma)$ the familiar $u$(N)-valued gauge fields and scalars of 1+1 dimensional super Yang Mills theory.  It is also useful to note a relationship between D1-brane integration and D0-brane trace:
\bea
\int d\sigma\ \tr\left(X(\sigma)\right) &=& \tr\left(X(0)\right) \nonumber\\
&=& {\tr(X) \over \tr(\mathbf{1})/N},\label{Tr}\\
&=& \Tr(X). \nonumber
\eea
The last line defines $\Tr$ to be the trace over only a single copy of the D0 $S^1$.

The T-dual picture is useful to gain intuition for BPS conditions of the matrix model.  We see that
\be
[Z,\Phi] &=& -iL\ds \Phi + [A_\sigma,\Phi]\\
&\equiv& \ \ L D_\sigma\Phi
\ee
acts as the covariant derivative along the dual string, so that the BPS conditions $D_0 X^i + i[Z,X^i]=0$ translate to $\omega=-k_\sigma$, the requirement that modes propagate in only one direction on the string.  Similarly, the condition
\be
D_0 Z=0 &\Longrightarrow& \mathbf{F}=0,
\ee
requiring that the D1 gauge connection is flat.

\section{$N=1$}

The simplest case is that of a single D0 on $S^1$, which nonetheless reveals a wealth of structure.  Since the adjoint representation of a U(1)-connection is trivial, the action upon $X^i$ of a possible Wilson loop in $A^\mu$ is trivial, and we can choose a gauge in which $A^\sigma=A^0=0$.  The remaining BPS conditions are then solved by 
\bea
X^i(\sigma, t)= \sum_{l} X^i(l)e^{il(\sigma-t)}. \label{Xt}
\eea
One can easily check that the Gauss-law constraint (\ref{Gauss}) is automatically satisfied, as any two matrices that can be written as a function of $\sigma$ commute.  The general $N=1$ solution to (\ref{BPS}) is then specified by the set of $X^i(n)$, with $Z_{mn}=mL\delta_{mn}$ and $A^0=0$.

\subsection{The Bloch Wave Basis}

Since $[X^i,X^j]=0$, we can make sense of these matrix configurations by diagonalizing the $X^i$.  To do so, we take advantage of an analogy between the form (\ref{X}) of these matrices and the Hamiltonian of a homogeneous, infinite, 1-dimensional crystal, due to their shared invariance under shifts $\mathbf{Z}$.  According to the analogy, $X^i(l)$ is understood as the coupling between a lattice site and its $l^{th}$ nearest neighbor.  We can then simply write down the Bloch wave eigenvector
\be
v_m = e^{ims},
\ee
for which the continuous parameter $s$ plays the role of the Bloch wave number.  Amusingly, the periodicity of wave number on the lattice $k \sim k+{2\pi \over b}$ is responsible for making $s$ a periodic coordinate, making the solutions tubular.  The eigenvalues of this vector are simply found:
\bea
(X^i v)_m &=& \sum_{l,n} X^i(l)\left(e^{il\sigma}\right)_{mn} v_n \nonumber\\
&=& \sum_{l,n} X^i(l)\delta_{m,n-l} v_n \nonumber\\
&=& \sum_{l} X^i(l)v_{m+l} \label{eigen}\\
&=& \left(\sum_{l} X^i(l) e^{ils}\right) v_{m}.\nonumber
\eea
We see that the eigenvalues of $X^i$ trace out a closed curve $X^i(s)=\left(\sum X^i(l) e^{ils}\right)$ in $\mathbf{R}^8$, whose $l^{th}$ Fourier coefficient is nothing other than the matrix data $X^i(l)$.  It is now a simple matter to construct the fuzzy cylinder of arbitrary cross-section in $\mathbf{R}^8$, by merely taking the Fourier transform of its coordinates.  There is nothing particularly mysterious about fuzzy cylinders, once the appropriate Bloch wave basis is found.  

It should be noted that while the matrices $X^i$ in eq. (\ref{Xt}) are time-dependent in this gauge, time evolution acts on the eigenvalue distributions, $X^i(s,t)=\sum X^i(l)e^{i(s-t)}$ merely as a relabeling, having no effect on the physical configuration.  A gauge transformation could eliminate the time-dependence altogether.

\subsection{Relation to the Solutions of Bak and Lee}

In \cite{BL}, Bak and Lee presented the axially symmetric fuzzy cylinders, which they found by solving the BPS and Gauss-law constraints with the algebra of the two-dimensional Euclidean group:
\bea
[X,Z]&=&-ilY,\nonumber\\
\left[Y,Z\right]&=&ilX,\label{E2}\\
\left[X,Y\right]&=&0.\nonumber
\eea
The Casimir operator $X^2+Y^2=\rho^2$ sets the radius of the cylinder.  We can easily recover these solutions, viewed as living on the full non-compact covering space, by solving the differential equations implied by (\ref{E2}):
\be
iL\ds X(\sigma) &=& -ilY(\sigma)\\
iL\ds Y(\sigma) &=& ilX(\sigma).
\ee 
These are recognized as the equations of a circle, upon identifying the non-commutativity parameter $l$ with $L$, the periodicity scale, giving the non-vanishing coefficients
\be
X(0)=x_0,& \hspace{.25 in} &X(1)=\rho,\\
 Y(0)=y_0,& \hspace{.25 in} &Y(1)=i\rho.
\ee

\section{Generalization to $N > 1$}
The Bloch wave diagonalization is easily generalized to the case of $N>1$ D0 branes.  The eigenvectors are $\mathbf{v}_m=\mathbf{v}_0 e^{ims}$, with $\mathbf{v}_0$ an $N$-dimensional eigenvector of the $N \times N$ matrix
\be
X^i(s)=\sum_l X^i(l) e^{il s}.
\ee
$N$ disconnected fuzzy cylinders are achieved by choosing the $X^i(l)=diag(x^i_{(1)}(l),\dots,x^i_{(N)}(l))$, with the $x^i_{(n)}(l)$ the Fourier coefficients of the individual profiles.

\subsection{Wilson Loops}
The diagonal case is not sufficiently general, however, as the U($N$) gauge theory on the T-dual D1's can admit Wilson loops which permute the segments of string upon winding the dual $S^1$.  Rather than setting $A^\sigma$ to zero, we must take into account its zero modes, $A^\sigma(0)\in S_N$, which can play the important role of piecing together the $N$ separate D1's as $\sigma \rightarrow \sigma +2\pi$, to arrive at $<N$ longer strands.

Since any element $\rho \in S_N$ satisfies $\rho^n=\mathbf{1}$ for some $n$, we can eliminate a non-trivial Wilson loop by going to the $n$-fold cover of the D1's $S^1$.\footnote{Note that the covering space of the D1 circle is not at all the same as the covering space of the D0 $S^1$.}  The Wilson loop linking $n$ D1 branes on a circle of size $1/L$ is equivalent to a trivial connection with a single D1 brane living on a circle of size $n/L$.  We can therefore construct the $N>1$ solutions by repackaging the matrices of the $N=1$ solutions on a circle of size ${L\over n}$

For example, the non-trivial Wilson loop for two D0's is the permutation $\rho=(12)$, which ties them into a single long loop.  The solution is nothing but the $N=1$ solution on $L/2$, repackaged in an $N=2$ form, schematically:
\be
Z&=&
\dots \left( \begin{array}{cc}

\left( \begin{array}{cc}
{L \over 2}&0\\
0&L\\
\end{array}\right) & 
\left( \begin{array}{cc}
0&0\\
0&0
\end{array}\right)
\\
\left( \begin{array}{cc}
0&0\\
0&0
\end{array}\right)
& \left( \begin{array}{cc}
{3L \over 2}&0\\
0&2L
\end{array}\right)

\end{array}\right) \dots \\
X&=&
\dots \left( \begin{array}{cc}

\left( \begin{array}{cc}
x(0)& x(1)\\
x(1)^\dagger&x(0)\\
\end{array}\right) & 
\left( \begin{array}{cc}
x(2)&x(3)\\
x(1)&x(2)
\end{array}\right)
\\
\left( \begin{array}{cc}
x(2)^\dagger &x(1)^\dagger\\
x(3)^\dagger &x(2)^\dagger
\end{array}\right)
& 
\left( \begin{array}{cc}
x(0)& x(1)\\
x(1)^\dagger&x(0)\\
\end{array}\right)

\end{array}\right)\dots
\ee
This is a state with half a unit of F-charge.

Solutions with more than one multiwound string can be built by taking $X^\mu(\sigma)$ to be block-diagonal, with the single-string solutions in each block.

\section{Charges of the Solutions}

If the solutions considered here are to be identified with supertubes, they must be shown to carry the fundamental string charge and angular momentum of supertubes.  It is again convenient to think in terms of the T-dual picture, in which F-charge is the momentum along the circle, $-i\ds$.  

That is,
\be
N_f &=& \int d\sigma\ T^1_{\ 0}
= \int d\sigma\ \tr(\p_0X^i \p_\sigma X_i) \\
&=& -\int d\sigma\ \tr (\p_\sigma X^i \p_\sigma X_i) 
= -\Tr(\p_0 X^i \p_0 X_i) \\
&=& -H.
\ee
We have used (\ref{Tr}) and (\ref{BPS}), to get a result obvious from the T-dual perspective, where F-charge is momentum: These D0 brane configurations carry F-charge proportional to their energy.  

The final ingredient for a supertube is angular momentum:
\be
J_{ij}&=&\Tr(X_iD_0X_j -X_jD_0 X_i)
=\Tr(X_i\p_\sigma X_j -X_jD_\sigma X_i)\\
&=&\int d\sigma\ \tr(X_i\p_\sigma X_j - X_j\p_\sigma X_i)
=2\int d\sigma\ \tr(X_i\p_\sigma X_j).
\ee
Taking $X_i(s) = diag(x_i^{(1)},\dots,x_i^{(N)})$, this becomes
\be
J_{ij}&=&2\sum_{n=1}^N \int dx_j^{(n)} x_i^{(n)}\\
&=& 2\sum_{n=1}^N A^{(n)}_{ij}.
\ee
The angular momentum in the $(i,j)$-plane is nothing but the area enclosed by the curves projected onto that plane.  It is intriguing that the area enclosed by the curve would be integer-quantized, despite the commutativity of the operators defining that plane.

\section{Conclusions}
With the aid of a Bloch wave basis, we have constructed a wide class of quarter-BPS objects in matrix theory: the non-commutative cylinders of arbitrary cross-section in $\mathbf{R}^8$.  The Bloch wave basis elucidates how the matrix degrees of freedom are reorganized into extended objects, with the eigenvalues of the transverse matrices tracing a smooth curve.  The solutions demonstrate a number of interesting features, such as fractional F-charge and area quantization, and provide a microscopic description of the most general supertube.

\section{Acknowledgments}

We would like to thank Eric Gimon and Andy Charmin for valuable discussion.  This material is based upon work supported in part by NSF grant PHY-0244900, by the Berkeley Center for Theoretical Physics, and by DOE grant DE-AC02-05CH11231.  

\bibliographystyle{board}
\bibliography{all}

\begin{thebibliography}{10}


\bibitem{Marty}
  M.~Claudson and M.~B.~Halpern,
  ``Supersymmetric Ground State Wave Functions,''
  Nucl.\ Phys.\ B {\bf 250}, 689 (1985).

\bibitem{BFSS}
  T.~Banks, W.~Fischler, S.~H.~Shenker and L.~Susskind,
  ``M theory as a matrix model: A conjecture,''
  Phys.\ Rev.\ D {\bf 55}, 5112 (1997)
  [arXiv:hep-th/9610043].

\bibitem{BSS}
  T.~Banks, N.~Seiberg and S.~H.~Shenker,
  ``Branes from matrices,''
  Nucl.\ Phys.\ B {\bf 490}, 91 (1997)
  [arXiv:hep-th/9612157].

\bibitem{Matrix}
  D.~Bigatti and L.~Susskind,
  ``Review of matrix theory,''
  [arXiv:hep-th/9712072].

  W.~Taylor,
  ``M(atrix) theory: Matrix quantum mechanics as a fundamental theory,''
  Rev.\ Mod.\ Phys.\  {\bf 73}, 419 (2001)
  [arXiv:hep-th/0101126].

\bibitem{Lexi}
  R.~Bousso and A.~L.~Mints,
  ``Decoding the matrix: Coincident membranes on the plane wave,''
  [arXiv:hep-th/0510121].

\bibitem{BL}
  D.~Bak and K.~M.~Lee,
  ``Noncommutative supersymmetric tubes,''
  Phys.\ Lett.\ B {\bf 509}, 168 (2001)
  [arXiv:hep-th/0103148].

\bibitem{LRS}
  D.~Loh, K.~Rudolfa and V.~Sahakian,
  ``Non-commutative dynamics of spinning D0 branes,''
  [arXiv:hep-th/0408072].

\bibitem{BOS}
  D.~s.~Bak, N.~Ohta and M.~M.~Sheikh-Jabbari,
  ``Supersymmetric brane anti-brane systems: Matrix model description,
  stability and decoupling limits,''
  JHEP {\bf 0209}, 048 (2002)
  [arXiv:hep-th/0205265].

\bibitem{HO}
  Y.~Hyakutake and N.~Ohta,
  ``Supertubes and supercurves from M-ribbons,''
  Phys.\ Lett.\ B {\bf 539}, 153 (2002)
  [arXiv:hep-th/0204161].

\bibitem{BK}
  D.~s.~Bak and A.~Karch,
  ``Supersymmetric brane-antibrane configurations,''
  Nucl.\ Phys.\ B {\bf 626}, 165 (2002)
  [arXiv:hep-th/0110039].

\bibitem{BO}
D.~s.~Bak and N.~Ohta,
``Supersymmetric D2 anti-D2 strings,''
Phys.\ Lett.\ B {\bf 527} (2002) 131
[arXiv:hep-th/0112034]



\bibitem{Supertubes}
  D.~Mateos and P.~K.~Townsend,
  ``Supertubes,''
  Phys.\ Rev.\ Lett.\  {\bf 87}, 011602 (2001)
  [arXiv:hep-th/0103030].

  D.~Mateos, S.~Ng and P.~K.~Townsend,
  ``Tachyons, supertubes and brane/anti-brane systems,''
  JHEP {\bf 0203}, 016 (2002)
  [arXiv:hep-th/0112054].

\bibitem{Myers}
  R.~C.~Myers,
  ``Dielectric-branes,''
  JHEP {\bf 9912}, 022 (1999)
  [arXiv:hep-th/9910053].

  R.~C.~Myers,
  ``Nonabelian D-branes and noncommutative geometry,''
  Int.\ J.\ Mod.\ Phys.\ A {\bf 16}, 956 (2001)
  [arXiv:hep-th/0106178].

\bibitem{Taylor}
  W.~I.~Taylor,
  ``D-brane field theory on compact spaces,''
  Phys.\ Lett.\ B {\bf 394}, 283 (1997)
  [arXiv:hep-th/9611042].


\end{thebibliography}
\end{document}